\documentclass[11pt]{article}
\usepackage{amssymb}
\usepackage{amssymb}
\usepackage{amsmath}
\usepackage{latexsym}
\usepackage{theorem}
\usepackage{graphicx}
\usepackage{graphics}
\usepackage{epsfig}
 at 12 truept
\begin{document}
\date{}
\title{Axiomatic field theory and Hida-Colombeau algebras}
\author{Alexei Filinkov\\
 School of Mathematical Sciences\\
University of Adelaide\\
Adelaide SA 5005, Australia\\
alexei.filinkov@adelaide.edu.au
\and Ian G. Fuss\\
School of Electrical and Electronic Engineering\\
University of Adelaide\\
Adelaide SA 5005, Australia\\
ian.fuss@adelaide.edu.au}
\maketitle
\date						% Activate to display a given date or no date

%\begin{document}
%\maketitle
\begin{abstract}
An axiomatic quantum field theory applied to the self interacting boson field is realised in terms of generalised operators that allows us to form products and take derivatives of the fields in simple and mathematically rigorous ways. Various spaces are explored for representation of these operators with this exploration culminating with a Hida-Colombeau algebra.

Rigorous well defined Hamiltonians are written using ordinary products of interacting scalar fields that are represented as generalised operators on simplified Hida-Colombeau algebras.
\end{abstract}
%\section{}
%\subsection{}
\section{Introduction}
 It is conventional to express quantum fields as operator-valued functions of space and time:
 \begin{equation}\label{eq1}
 \phi = \phi(t,\mathbf{x}) = \phi(x)\,, \ \mathrm{where}\ x = (t, \mathbf{x})\in \mathbb{R}^4\,.
 \end{equation}
Within the framework of relativistic quantum theory \cite{Coleman} these operators need to satisfy the locality condition, be self-adjoint, satisfy the translation and scalar Lorentz transformation properties.

In analogy with classical field theory the simplest operators  $\phi$ satisfying these conditions are expressed/constructed in terms of annihilation $a_\mathbf{p}$ and creation $a_\mathbf{p}^\dag$ operators, where $\mathbf{p}\in \mathbb{R}^3$ is a three momenta of a particle with mass $m>0$.  The reduction from the space-time space $\mathbb{R}^4$ of the equation (\ref{eq1}) to the three-momentum space $\mathbb{R}^3$ here is a result of a constraint that is imposed by Lorentz symmetry \cite{Coleman}. 

Quantum field theory is a many body description of nature. In order to facilitate this we adopt the convention of defining the operators $\phi$ on the boson Fock space \cite{Folland}
 \[
 \Gamma (\mathcal{H}) = \oplus_{n=0}^{\infty} \mathcal{H}^{\widehat{\otimes}n}\equiv  \oplus_{n=0}^{\infty}\Gamma_n (\mathcal{H})\,,
 \]
the direct sum of symmetric tensor algebras $\mathcal{H}^{\widehat{\otimes}n}$ of a Hilbert space $\mathcal{H} = L^2(\mathbb{R}^3, d\sigma(\mathbf{p}))$ of relativistic spin zero single particles.

However,  creation and annihilation operators, and consequently scalar field operators, are unbounded  on the Fock space $ \Gamma (\mathcal{H})$ and they are usually regarded as operator-valued distributions \cite{Wigh}.

The operator-valued distribution formalism permits the expression of equal time canonical commutation relations for the field $\phi$ and its canonically conjugate field $\pi (x)$ \cite{Folland}.

However, the system's Hamiltonian operator-valued distribution $H$ is ill-defined as is contains terms that are at least quadratic in $\phi$. Schwartz's impossibility of multiplication of distributions theorem \cite{Schwartz} raises significant questions in relation to the meaning of such operators. It is customary to invoke Wick (normal) ordering of creation and annihilation operator-valued distributions in order to give mathematical meaning to these expressions \cite{Folland}.

In this note we will show that if the field $\phi$ is treated as a Colombeau-type generalised operator on simplified Hida-Columbeau algebras, the complexity of Wick-ordered products and the associated obfuscation  of the dynamic models is removed since the field products can be meaningfully defined in such setting in terms of multiplication of $C^\infty$-functions without invoking the artifice of  removing divergences by hand.

%%%%%%%%%%%%%%%%%%%%%%%%%%%%%%%%%%%%%%%%%%%%%%%%%%%%%
 \section{Fock Space}

Following convention, we now define the boson Fock space
 \[
 \Gamma (\mathcal{H}) = \oplus_{n=0}^{\infty} \mathcal{H}^{\widehat{\otimes}n}\equiv  \oplus_{n=0}^{\infty}\Gamma_n (\mathcal{H})
 \]
as the direct sum of symmetric tensor algebras $\mathcal{H}^{\widehat{\otimes}n}$ of the Hilbert space $\mathcal{H} = L^2(\mathbb{R}^3, d\sigma(\mathbf{p}))$ of relativistic spin zero single particles, where $d\sigma(\mathbf{p}) = d\mathbf{p}/\omega(\mathbf{p})$ with  $\omega(\mathbf{p})\equiv p^0(\mathbf{p}) = \sqrt{\vert\mathbf{p} \vert^2 + m^2} $, is a Lorentz invariant measure on $\mathbb{R}^3$.

For $f\in \mathcal{H}$  we define the creation operator on the boson Fock space
\[
A^\dag (f):  \Gamma (\mathcal{H}) \to  \Gamma (\mathcal{H})
\]
by the rule
\[
A^\dag (f) F_n = \sqrt{n+1}\, f \hat{\otimes} F_n\,,
\]
where 
\[
F=(F_n)_{n=0}^\infty\in \Gamma (\mathcal{H})\quad\mathrm{with} \quad F_n\in \Gamma_n (\mathcal{H})\,,\quad\mathrm{and} \quad f \hat{\otimes} F_n\in \Gamma_{n+1} (\mathcal{H})\,.
\]
The annihilation operator $A(f):  \Gamma (\mathcal{H}) \to  \Gamma (\mathcal{H})$ is the adjoint to $A^\dag (f)$, it corresponds to the rule
\[
A (f) F_n = \frac{1}{\sqrt{n}}\,\sum_{j=1}^n  (f, F_n^j )_{\mathcal{H}} \, F_n^1\hat{\otimes} \ldots\hat{\otimes} \breve{F_n^j}\hat{\otimes} \ldots\hat{\otimes} F_n^n\in \Gamma_{n-1} (\mathcal{H})\,,
\]
where $(f, F_n^j )_{\mathcal{H}}$ is inner product in $\mathcal{H}$,  $ F_n = F_n^1\hat{\otimes} \ldots\hat{\otimes}  F_n^n$ and $\breve{F_n^j}$ is omitted. It has the property $A(f)\Gamma_0 (\mathcal{H}) = 0$. 

Application of operator  $A^\dag (f)$ can be interpreted as addition of a particle in state $f$ to a multi-particle state $F_n$, whilst application of  $A (f)$ as removal of a particle in state $f$ from a multi-particle state $F_n$. In order to add or remove particles of momentum $\mathbf{p}$, we extend these operators to the boson Fock-Gelfand triple
\[
 \Gamma \big(\mathcal{S}(\mathbb{R}^3)\big) \subset  \Gamma (\mathcal{H}) \subset  \Gamma \big(\mathcal{S}'(\mathbb{R}^3)\big) \,,
\]
where $ \Gamma \big(\mathcal{S}(\mathbb{R}^3)\big) = \oplus_{n=0}^{\infty} \mathcal{S}(\mathbb{R}^3)^{\widehat{\otimes}n}$. it is based on the Gelfand triple
\[
 S(\mathbb{R}^3) \subset \mathcal{H} \subset S'(\mathbb{R}^3)  \, ,
\]
where $S(\mathbb{R}^3)$ is the Schwartz space of rapidly decreasing $C^\infty$ functions and  $S'(\mathbb{R}^3)$ is the space of Schwartz tempered distributions. We note that operators $A(f)$ and $A^\dag (f)$ are bounded on $ \Gamma \big(\mathcal{S}(\mathbb{R}^3)\big)$.

We now define generalised operators
\[
a_ \mathbf{p}^\dag := A^\dag (\delta_ \mathbf{p}(\mathbf{k}))\quad \mathrm{and} \quad a_ \mathbf{p} := A (\delta_ \mathbf{p}(\mathbf{k}))\,,
\]
where $\delta_ \mathbf{p}(\mathbf{k}) = \delta (\mathbf{k} - \mathbf{p})\in \mathcal{S}^\prime(\mathbb{R}^3)$ is the Dirac measure. Thus, we have the generalised annihilation operator $a_ \mathbf{p}:  \Gamma \big(\mathcal{S}(\mathbb{R}^3)\big) \to  \Gamma (\mathcal{H})$ with the rule
\begin{equation}\label{anoper}
a_ \mathbf{p} F_n = \frac{1}{\sqrt{n}}\,\sum_{j=1}^n  \langle \delta_ \mathbf{p}(\mathbf{k}), F_n^j \rangle \, F_n^1\hat{\otimes} \ldots\hat{\otimes} \breve{F_n^j}\hat{\otimes} \ldots\hat{\otimes} F_n^n\in \Gamma_{n-1} (\mathcal{H})\,,
\end{equation}
with the dual pair $\langle \delta_ \mathbf{p}(\mathbf{k}), F_n^j (\mathbf{k})\rangle = F_n^j (\mathbf{p})$ \cite{Vladimirov},  and generalised creation operator $a_ \mathbf{p}^\dag : \Gamma \big(\mathcal{S}(\mathbb{R}^3)\big) \to  \Gamma \big(\mathcal{S}^\prime(\mathbb{R}^3)\big) $ with the rule
\begin{equation}\label{croper} 
a_ \mathbf{p}^\dag F_n = \sqrt{n+1}\, \delta_ \mathbf{p}(\mathbf{k}) \hat{\otimes} F_n\in  \mathcal{S}^\prime(\mathbb{R}^3)^{\widehat{\otimes}(n+1)}\,.
\end{equation}
Here formulae (\ref{anoper}) and  (\ref{croper}) are consistent with the previously mentioned interpretation that operator $a_ \mathbf{p}$ removes particles of momentum $\mathbf{p}$ and operator $a_ \mathbf{p}^\dag$ adds them.

%%%%%%%%%%%%%%%%%%%%%%%%%%%%%%
\section{Scalar Quantum Fields}
 We will build a scalar quantum field as an operator-valued "function" of space and time:
 \[
 \phi = \phi(t,\mathbf{x}) = \phi(x)\,, \ \mathrm{where}\ x = (t, \mathbf{x})\,.
 \] 
\subsection{Axioms}
Within the framework of relativistic quantum theory \cite{Coleman} these operators need to 
\begin{enumerate}
\item satisfy the locality condition
\[
[\phi(x), \phi(y)] = 0\ \mathrm{for\, any} \ x, y\in \mathbb{R}^4\ \mathrm{such\, that} \ (x-y)^2 < 0\,;
\]
\item be self-adjoint
\[
\phi(x) = \phi(x)^\dagger \ \mathrm{for\, any} \ x\in \mathbb{R}^4\,;\ \mathrm{and}
\]
\item satisfy the translation
\[
e^{-i P\cdot y} \phi(x) e^{i P\cdot y}= \phi(x-y) \ \mathrm{for\, any} \ x, y\in \mathbb{R}^4
\]
and scalar Lorentz transformation 
\[
U(\Lambda)^\dag  \phi(x) U(\Lambda) = \phi(\Lambda^{-1} x)\ \mathrm{for\, any} \ x\in \mathbb{R}^4
\]
properties.
\end{enumerate}

Here $P = (H, \mathbf{P})$ is the four-momentum operator, where $H$ is the Hamiltonian and $\mathbf{P}$ is the three-momentum operator on $\mathcal{H}$. Operator $P$ is the generator of a space-time translation group and $e^{i P\cdot y}$ is a space-time translation operator for a state $\psi\in \mathcal{H}$. Further, $\Lambda$ is a scalar proper orthochronous Lorentz transformation matrix and $ U(\Lambda)$ is a scalar Lorentz transformation operator for a state $\psi\in \mathcal{H}$.

\subsection{Axiomatic Fields}
In analogy with classical field theory it is natural to start to search for operators  $\phi$ satisfying conditions 1-3 as linear combinations of creation and annihilation operators:
\begin{equation}\label{lincomb}
\phi(x) = \int_{\mathbb{R}^3} d\mathbf{p} \Big( f_\mathbf{p} (x) a_\mathbf{p} + g_\mathbf{p}(x) a_\mathbf{p}^\dag\Big)\,,
\end{equation}
where $f_\mathbf{p}$ and $g_\mathbf{p}$ are some coefficients depending on $x$.

The transformation properties of creation and annihilation operators
\[
e^{i P\cdot x} a_\mathbf{p} e^{- i P\cdot x} = e^{- i P\cdot x} a_\mathbf{p}\,,\quad e^{i P\cdot x} a_\mathbf{p}^\dag e^{- i P\cdot x} = e^{ i P\cdot x} a_\mathbf{p}^\dag
\]
and
\[
U(\Lambda)  a_\mathbf{p} U(\Lambda)^\dag = a_\mathbf{\Lambda p}\,,\quad U(\Lambda)  a_\mathbf{p}^\dag U(\Lambda)^\dag = a_\mathbf{\Lambda p}^\dag
\]
imply that the only operators of type (\ref{lincomb}) that satisfy conditions 1-3 has are generalised operators of the form
\begin{equation}\label{lincomb2}
\phi(x) = \int_{\mathbb{R}^3} \frac{d\mathbf{p}}{\sqrt{\omega(\mathbf{p})}} \Big( e^{-i p\cdot x} a_\mathbf{p} + e^{i p\cdot x} a_\mathbf{p}^\dag\Big)\,
\end{equation}
Differentiating twice with respect to $x$ implies that these generalised operators satisfy the Klein-Gordon equation \cite{Coleman}
\begin{equation}\label{KG}
\square \phi(x) + m \phi(x) = 0\,,
\end{equation}
where\footnote{here $\partial_0 = \partial/\partial t $}  $\square = \partial^2_0 - \bigtriangledown_\mathbf{x}^2$. We can also write 
\[
\phi(x) = \int_{\mathbb{R}^3} \frac{d\mathbf{p}}{\sqrt{\omega(\mathbf{p})}} \Big( e^{-i p\cdot x} a_\mathbf{p} + e^{i p\cdot x} a_\mathbf{p}^\dag\Big) \equiv \phi_-(x) + \phi_+(x)\,,
\]
where
\begin{equation}\label{lincomb22}
 \phi_-(x) = \int_{\mathbb{R}^3} \frac{d\mathbf{p}}{\sqrt{\omega(\mathbf{p})}} e^{-i p\cdot x} a_\mathbf{p}\quad\mathrm{and}\quad  \phi_+(x) = \int_{\mathbb{R}^3} \frac{d\mathbf{p}}{\sqrt{\omega(\mathbf{p})}}  e^{i p\cdot x} a_\mathbf{p}^\dag \,,
\end{equation}
which we will refer to as annihilation and creation fields, respectively.

%%%%%%%%%%%%%%%%%%%%%%%%%%%%%%%%%%%%%%
\section{Fock space and Hida generalised functions}

In order to be able to analyse products of fields, creation and annihilation operators which are represented as generalised operators, we now consider a specific realisation of the boson Fock space that was introduced by Hida \cite{Hida, Obata,Kuo}. Following the convention we denote  $E := \mathcal{S} (\mathbb{R}^3)$, $ E^\ast := \mathcal{S}^\prime (\mathbb{R}^3)$ and  $(L^2) := L^2 (\mathcal{S}^\prime (\mathbb{R}^3), \mu)$, where  $\mu$ is a Gaussian measure.
Note the Wiener-It$\hat{\mathrm{o}}$-Segal isomorphism 
\[
 \Gamma (\mathcal{H}) \cong (L^2)\,,
\]
where 
\[
f \in (L^2) \iff \exists (f_n)_{n=0}^{\infty} \in \Gamma (\mathcal{H}) = \oplus_{n=0}^{\infty} \mathcal{H}^{\widehat{\otimes}n}\,
\]
such that 
\[
f (\zeta) = \sum_{n=0}^{\infty} \langle :\zeta^{\otimes n}: \, , f_n\rangle\,, \qquad \zeta\in E^\ast ,
\]
where $\langle :\zeta^{\otimes n}: \, , f_n\rangle$ denotes a Wick polynomial \cite{Obata}. The inner product in this space has the form
\[
\langle f^1, f^2 \rangle_{(L^2)} = \sum_{n=0}^{\infty} n!\  \langle f_n^1, f_n^2 \rangle_{\Gamma_n (\mathcal{H})}\,,
\]
where $f^1, f^2\in (L^2) $ and $(f_n^1)_{n=0}^{\infty}, (f_n^2)_{n=0}^{\infty} \in \Gamma (\mathcal{H})$.

 Consider the Gelfand-Hida triple
\[
(E) \subset (L^2)\subset (E)^\ast\,,
\]
where the Hida space of test functions $(E)$ is  characterised in terms of the Wiener-It$\hat{\mathrm{o}}$-Segal isomorphism: 
\[
f \in (E) \iff \exists  (f_n)_{n=0}^{\infty} \in \Gamma (E) = \oplus_{n=0}^{\infty} E^{\widehat{\otimes} n}\,,
\]
the Fock space over $E$, with $f_n \in E^{\widehat{\otimes}  n}$ for each $n$ and 
\begin{equation}\label{testf}
f (\zeta) = \sum_{n=0}^{\infty} \langle :\zeta^{\otimes n}: \, , f_n\rangle\,, \qquad \zeta\in E^\ast\,.
\end{equation}
Likewise the Hida space of generalised functions is  characterised in terms of 
\[
F \in (E)^\ast \iff \exists (F_n)_{n=0}^{\infty} \in \Gamma (E)^\ast = \oplus_{n=0}^{\infty} (E^{\widehat{\otimes} n})^\ast\,,
\]
with $F_n \in  (E^{\widehat{\otimes}  n})^\ast$ for each $n$ and  formal expressions
\[
F (\zeta) = \sum_{n=0}^{\infty} \langle :\zeta^{\otimes n}: \, ,F_n\rangle\,, \qquad \zeta\in  E^\ast .
\]
The dual pairing of $F \in (E)^\ast $ and $f \in (E)$ has the form
\[
 \langle\langle F\,, f\rangle\rangle = \sum_{n=0}^{\infty} n!  \langle F_n \, , f_n\rangle\,.
\]

%%%%%%%%%%%%%%%%%%%%%%%%%%%%%%%%%%%%%%%%%%%%
\section{Annihilation and creation fields as generalised operators in Hida space}
%\subsection{Annihilation fields}
For $f \in (E)$ of the form (\ref{testf}), we consider the operator
\[
D_z \, f(\zeta) = \sum_{n=0}^{\infty} n \langle :\zeta^{\otimes (n-1)}: \, , z \, \hat{\otimes}_1 \,  f_n \rangle\,, \qquad \zeta,z \in E^\ast ,
\]
where $z \, \hat{\otimes}_1 \,  f_n \in E^{\hat{\otimes} (n-1)}$ is a symmetric contraction \cite{Obata, Huang}. For any $z\in E^\ast$, the operator $D_z\in \mathcal{L}\big((E), (E)\big)$ is a continuous derivation on $(E)$ \cite{Obata}:
\[
D_z (f_1 \,f_2) = (D_z f_1)\,f_2 + f_1 (D_z f_2),\quad\quad  f_1 \,,f_2\in (E)\,,
\]
and $D_z \, f(\zeta)$ is the Gateaux differential of $f$ at $\zeta\in E^\ast$ in the direction $z \in E^\ast$:
\[
D_z \,  f(\zeta) = \lim_{\epsilon \rightarrow 0} \frac{f(\zeta + \epsilon z) - f(\zeta)}{\epsilon} \,.
\]

In particular, if $z = \delta_{\mathbf{p}}$, then $D_{\delta_{\mathbf{p}}} := \partial_{\mathbf{p}}$ is called Hida's differential operator, and we have
\[
\partial_{\mathbf{p}} f(\zeta) = \lim_{\epsilon \rightarrow 0} \frac{f(\zeta + \epsilon \delta_{\mathbf{p}}) - f(\zeta)}{\epsilon} \,,
\]
and is a Hida space representation of the annihilation operator $a_\mathbf{p}$ (\ref{anoper}).
This enables us to write a Hida space representation of the annihilation field  $\phi_- $ (\ref{lincomb22}) in the form
\[
\Phi_- (t, \mathbf{x})  = \int_{\mathbb{R}^3} \frac{d\mathbf{p}}{\sqrt{\omega(\mathbf{p})}} \, e^{-i(\omega_\mathbf{p}t -\mathbf{p} \cdot \mathbf{x})} \partial_\mathbf{p} \equiv  \int_{\mathbb{R}^3} d\mathbf{p} \, z_1(\mathbf{p}) \partial_\mathbf{p} = D_{z_1} (t, \mathbf{x}) \, ,
\]
where $z_1(\mathbf{p}) =  \frac{1}{\sqrt{\omega(\mathbf{p})}}  \, e^{-i(\omega_\mathbf{p}t -\mathbf{p} \cdot \mathbf{x})} \,$.

%\subsection{The canonical momentum of the annihilation field}
It is the usual practice in quantum field theory to define the canonical momentum field
\[
\Pi (t, \mathbf{x}) = \frac{\partial}{\partial t} \Phi(t, \mathbf{x})
\]
for the field $\Phi$ \cite{Mag}. Hence we write a Hida space representation of the canonical momentum field
\begin{equation} \label{amf}
\Pi_-(t, \mathbf{x}) = \frac{\partial \,\,}{\partial t} \int_{\mathbb{R}^3} d\mathbf{p} \,  z_1(\mathbf{p}) \partial_\mathbf{p} = -i\, \int_{\mathbb{R}^3} d\mathbf{p}\,\sqrt{\omega(\mathbf{p})} \, e^{-i(\omega_\mathbf{p}t -\mathbf{p} \cdot \mathbf{x})} \partial_\mathbf{p} \equiv -i\,D_{z_2}(t, \mathbf{x}) \, ,
\end{equation}
where $z_2(\mathbf{p}) = \sqrt{\omega(\mathbf{p})}  \, e^{-i(\omega_\mathbf{p}t -\mathbf{p} \cdot \mathbf{x})} \,$.

%\subsection{Creation fields}
The adjoint $D_z^*\in \mathcal{L}\big((E)^*, (E)^*\big)$ of the differential operator $D_z$ can be written in the form
\[
D_z^* \, f(\zeta) = \sum_{n=0}^{\infty} \langle :\zeta^{\otimes (n+1)}: \, , z \, \hat{\otimes} \,  f_n \rangle\,, \qquad \zeta,z \in E^\ast ,
\]
where $z \, \hat{\otimes} \,  f_n \in E^{\hat{\otimes} (n+1)}$. If $z = \delta_{\mathbf{p}}$, then $D_{\delta_{\mathbf{p}}}^* := \partial_{\mathbf{p}}^* $ is the dual of the Hida derivative, where by integration by parts we have
\[
\partial_{\mathbf{p}}^* f(\zeta) = [-\partial_{\mathbf{p}} +\zeta] f(\zeta) \, ,
\]
and $\partial_{\mathbf{p}}^*$ is a Hida space representation of the creation operator $a^*_\mathbf{p}$ (\ref{croper}).
The existence of this dual partial derivative enables us to write a Hida space representation of the creation field  $\phi_+$ (\ref{lincomb22}) in the form
\[
\Phi_+(t, \mathbf{x}) = \int_{\mathbb{R}^3} d\mathbf{p} \,\frac{1}{\sqrt{\omega(\mathbf{p})}} \, e^{i(\omega_\mathbf{p}t -\mathbf{p} \cdot \mathbf{x})}\partial_{\mathbf{p}}^* \equiv \int_{\mathbb{R}^3} d\mathbf{p} \, z_3(\mathbf{p}) \partial_{\mathbf{p}}^* = D_{z_3}^*(t, \mathbf{x})\, ,
\]
 where $z_3(\mathbf{p}) =  \frac{1}{\sqrt{\omega(\mathbf{p})}} \, e^{i(\omega_\mathbf{p}t -\mathbf{p} \cdot \mathbf{x})} \,$.

%\subsection{The canonical momentum of the creation field}
In a similar way to the canonical momentum of the annihilation field (\ref{amf}) we write a Hida space representation of the canonical momentum of the creation field as
\[
\Pi_+(t, \mathbf{x}) = \frac{\partial \,\,}{\partial t} \int_{\mathbb{R}^3} d\mathbf{p} \, z_3(\mathbf{p}) \partial_{\mathbf{p}}^* =  i\, \int_{\mathbb{R}^3} d\mathbf{p}\,\sqrt{\omega(\mathbf{p})} \, e^{i(\omega_\mathbf{p}t -\mathbf{p} \cdot \mathbf{x})} \partial_\mathbf{p} \equiv  i\,D_{z_4}^*(t, \mathbf{x}) \, ,
\]
where $z_4(\mathbf{p}) = \sqrt{\omega(\mathbf{p})}\, e^{i(\omega_\mathbf{p}t -\mathbf{p} \cdot \mathbf{x})} $.

%%%%%%%%%%%%%%%%%%%%%%%%%%%%%%%%%%%%%%%5
%\subsection{A commutator}
Thus a Hida space representation of the free real scalar field can be written in the form
\[
\Phi(t, \mathbf{x}) = \Phi_-(t, \mathbf{x}) + \Phi_+(t, \mathbf{x}) = D_{z_1}(t, \mathbf{x}) + D^*_{z_3}(t, \mathbf{x})\in \mathcal{L}\big((E), (E)^*\big) \, ,
\] 
and its canonical momentum field 
\[
\Pi(t, \mathbf{x}) = \Pi_-(t, \mathbf{x}) + \Pi_+(t, \mathbf{x}) = i\,(-D_{z_2}(t, \mathbf{x}) + D^*_{z_4}(t, \mathbf{x}))\in \mathcal{L}\big((E), (E)^*\big) \, ,
\]
that is $\Phi$ and $\Pi$ are generalised operators.
 The equal time canonical commutation relations for these fields
\[
 [\Phi(t, \mathbf{x_1}), \Phi(t, \mathbf{x_2})] = \Phi(t, \mathbf{x_1}) \Phi(t, \mathbf{x_2}) -\Phi(t, \mathbf{x_2}) \Phi(t, \mathbf{x_1}) \,,
\]
\[
 [\Pi(t, \mathbf{x_1}), \Pi(t, \mathbf{x_2})] = \Pi(t, \mathbf{x_1}) \Pi(t, \mathbf{x_2}) -\Pi(t, \mathbf{x_2}) \Pi(t, \mathbf{x_1}) \, ,
\]
\[
 [\Pi(t, \mathbf{x_1}), \Phi(t, \mathbf{x_2})] = \Pi(t, \mathbf{x_1}) \Phi(t, \mathbf{x_2}) -\Phi(t, \mathbf{x_2}) \Pi(t, \mathbf{x_1}) \, ,
\]
involve compositions of generalised differential operators: $D_\xi \, D_\eta$, $D_\xi \, D^*_\eta$, $D^*_\xi \, D_\eta$ and $D^*_\xi \, D^*_\eta$.
%%%%%%%%%%%%%%%%%%%%%%%%%%%%%%
 %\subsection{Compositions of generalised differential operators}

 Definitions of $D_z$ and  $D^*_z$ straightforwardly imply that  for all $\xi, \eta \in E^\ast$ compositions $D_\xi \, D_\eta$ and $D^*_\xi \, D^*_\eta$ are well-defined on $(E)$ and $(E)^* $, respectively, and we have 
\[
[D_\xi \,, D_\eta] = [D^*_\xi \,, D^*_\eta] = 0.
\]

The composition
 \[
 D^*_\xi \, [D_\eta \, f(\zeta)] = \sum_{n=0}^{\infty} n \langle :\zeta^{\otimes n}: \, , \xi \hat{\otimes}(\eta \, \hat{\otimes}_1 \,  f_n) \rangle\,, \qquad \zeta, \eta, \xi \in E^\ast 
 \]
is well-defined for $f\in (E)$.

 We note that the mapping
 \[
 D^*_\xi \, : \, (E)^* \, \rightarrow \, (E)^* \, ,
 \]
 is  well defined. In contrast, for $F\in  (E)^*$, the composition 
\[
D_\eta \, [D^*_\xi F(\zeta)]  = \sum_{n=0}^{\infty} (n + 1) \langle :\zeta^{\otimes n}: \, , \eta \hat{\otimes}_1(\xi \, \hat{\otimes} \,  F_n) \rangle\,,
\]
in general, is not well defined in $(E)^*$ for all $\xi, \eta \in E^\ast$ as it involves informal expressions like $\langle \eta\,, \xi\rangle$ and $\langle \eta\,, F_n\rangle$. 

However, these expressions are meaningful for functions $z_1 , ... , z_4$, and we have the following equal time canonical commutation relations
\[
 [\Phi(t, \mathbf{x_1}), \Phi(t, \mathbf{x_2})] = [\Pi(t, \mathbf{x_1}), \Pi(t, \mathbf{x_2})] = 0\, ,
\]
\[
 [\Pi(t, \mathbf{x_1}), \Phi(t, \mathbf{x_2})] = i\,\delta (\mathbf{x_1} - \mathbf{x_2})\,I\,
\]
in $\mathcal{L}\big((E), (E)^*\big)$.

Heisenberg and Pauli suggested  the Hamiltonian formulation in order to write a theory of interacting quantum fields \cite{Weinberg V1}. Initially we formally write a Hamiltonian for free scalar fields as
\begin{equation} \label{H1}
\mathbf{H(\phi)} = \int_{\mathbb{R}^3} H_0(\phi) \, d \mathbf{x} \, 
\end{equation}
and the free field Hamiltonian density \cite{Weinberg V1} can be wrtitten as
\begin{equation} \label{H3}
H_0(\phi) = \pi\dot{\phi}  - \mathcal{L}_0 = \frac{1}{2}[\pi^2 + (\nabla \phi)^2 + m^2 \phi^2 ] \, ,
\end{equation}
where we used the Legendre transform and the Lagrangian density is
\begin{equation} \label{L1}
\mathcal{L}_0(\phi) = \partial_\mu \phi^\dag \partial^\mu \phi -m^2 \phi^\dag \phi \, ,
\end{equation}
which has the Klein Gordon equation  (\ref{KG}) as its Euler Lagrange equation.

We have represented the scalar
quantum field by using generalised operators with domains and ranges that contain Schwartz
distributions. In his so called impossibility theorem Laurent Schwartz demonstrated
that the multiplication of these distributions does not always commute and their product may not
be a Schwartz distribution. In classical Quantum Field approaches these products are traditionally interpreted as Wick-ordered or normally-ordered products \cite{Folland, Huang, Luo}:
\[
H_0(\phi) =  \frac{1}{2}\, :[\pi^2 + (\nabla \phi)^2 + m^2 \phi^2 ]: \, ,
\]
which involve an unjustified removal of an infinite constant.

However Colombeau discovered algebras of generalised
functions that contain all the Schwartz distributions.
In the following we examine some of the technical details of a particular Colombeau
algebra that we will use to construct a new representation of quantum fields and
demonstrate that this representation simplifies the multiplication of scalar quantum
fields and hence scalar quantum field theory. Note that transition from the space of Hida distributions to a Hida-Colombeau algebra removes the necessity for the use of Wick-ordered products.

%\section{Simplified Colombeau algebras}
 \section{Hida-Colombeau algebras}
%\subsection*{Colombeau algebra of rapidly decreasing generalised functions}

We will introduce Hida-Colombeau algebras based on simplified Colombeau algebras \cite{Gros, Gsponer, Col}.
Let $\Omega$ be an open subset of $\mathbb{R}^3$. Consider a smooth function $f\in C^\infty (\Omega)$ and denote
\[
\mu_{q,l}(f) :=\sup_{x\in\Omega,\, \vert\alpha\vert\leq l}\big(1 + \vert x\vert\big)^q\,\big\vert \partial^\alpha f(x)\big\vert\,,\quad\mathrm{where}\ q\in \mathbb{Z}\ \mathrm{and}\ l\in\mathbb{N}\,.
\]
The set
\[
\mathcal{E}_\mathcal{S}(\Omega) := \Big\{ (f_\varepsilon)_\varepsilon\in \mathcal{S}(\Omega)^{(0,1]}\quad \mathrm{such\ that}\quad \forall q\,,l\in \mathbb{N}\  \exists n\in \mathbb{N}\,:\ \mu_{q,l}(f_\varepsilon) = O(\varepsilon^{-n})\ \mathrm{as}\ \varepsilon\to 0\Big\}
\]
is a sub-algebra of $\mathcal{S}(\Omega)^{(0,1]}$, where $\mathcal{S}(\Omega)$ is the Schwartz space of rapidly decreasing functions and
\[
\mathcal{S}(\Omega)^{(0,1]} = \big\{ (f_\varepsilon)_\varepsilon \ \mathrm{such\ that}\ \forall\varepsilon\in (0,1]\  f_\varepsilon\in \mathcal{S}(\Omega) \big\}
\]
is the set of nets. The algebraic set
\[
\mathcal{N}_\mathcal{S}(\Omega) := \Big\{ (f_\varepsilon)_\varepsilon\in \mathcal{S}(\Omega)^{(0,1]}\quad \mathrm{such\ that}\quad \forall q\,,l\in \mathbb{N}\  \forall p\in \mathbb{N}\,:\ \mu_{q,l}(f_\varepsilon) = O(\varepsilon^{p})\ \mathrm{as}\ \varepsilon\to 0\Big\}
\]
is an ideal in $\mathcal{E}_\mathcal{S}(\Omega)$. The factor-algebra
\[
\mathcal{G}_\mathcal{S}(\Omega) := \mathcal{E}_\mathcal{S}(\Omega)/\mathcal{N}_\mathcal{S}(\Omega)
\]
is referred to as the Colombeau algebra of rapidly decreasing generalised functions (see \cite{Del2007} and references therein).

%\subsubsection*{Colombeau algebras of tempered generalised functions}
Similarly, the factor-algebra
\[
\mathcal{G}_\tau(\Omega) := \mathcal{E}_\tau(\Omega)/\mathcal{N}_\tau(\Omega)
\]
is referred to as the Colombeau algebra of tempered generalised functions \cite{Colomb84}. Here
\[
\mathcal{E}_\tau(\Omega) := \Big\{ (f_\varepsilon)_\varepsilon\in \mathcal{O}_M(\Omega)^{(0,1]}\quad \mathrm{such\ that}\quad \forall l\in \mathbb{N}\  \exists q,\,n\in \mathbb{N}\,:\ \mu_{-q,l}(f_\varepsilon) = O(\varepsilon^{-n})\ \mathrm{as}\ \varepsilon\to 0\Big\}
\]
is a sub-algebra of $\mathcal{O}(\Omega)^{(0,1]}$, where
\[
\mathcal{O}_M(\Omega) := \Big\{ f\in C^\infty(\Omega)\ \mathrm{such\ that}\ \forall l\in \mathbb{N}\  \exists q\in \mathbb{N}\,:\ \mu_{-q,l}(f_\varepsilon) < \infty\Big\}
\]
is the algebra of smooth functions with slow growth (also known as the algebra of multiplicators); and the algebra
\[
\mathcal{N}_\tau(\Omega) := \Big\{ (f_\varepsilon)_\varepsilon\in \mathcal{O}_M(\Omega)^{(0,1]}\quad \mathrm{such\ that}\quad \forall l\in \mathbb{N}\ \exists q\in \mathbb{N}\ \forall p\in \mathbb{N}\,:\ \mu_{-q,l}(f_\varepsilon) = O(\varepsilon^{p})\ \mathrm{as}\ \varepsilon\to 0\Big\}
\]
is an ideal in $\mathcal{E}_\tau(\Omega)$. 
 %\section{Hida-Colombeau algebras}

We now define Hida-Colombeau algebra of rapidly decreasing generalised functions
\[
\mathcal{G}_\mathcal{S} (E) = \Gamma (\mathcal{G}_\mathcal{S} (\mathbb{R}^3)) =  \oplus_{n=0}^{\infty} \mathcal{G}_\mathcal{S} (\mathbb{R}^3)^{\widehat{\otimes}n}\,
\]
and Hida-Colombeau algebra of tempered generalised functions
\[
\mathcal{G}_\tau (E)^* = \Gamma (\mathcal{G}_\tau (\mathbb{R}^3)) =  \oplus_{n=0}^{\infty} \mathcal{G}_\tau (\mathbb{R}^3)^{\widehat{\otimes}n}\,.
\]
We have the embeddings
\[
 (E) \hookrightarrow\mathcal{G}_\mathcal{S} (E)\quad\mathrm{and}\quad (E)^*\hookrightarrow\mathcal{G}_\tau (E)^*\,,
\]
since for $f\in  (E)$ we have
\[
 f(\zeta) \rightarrow  \mathbf{f}(\zeta) = \Bigg( \sum_{n=0}^\infty \langle :\, \zeta^{\otimes n} \, : , f_n^\varepsilon \rangle \Bigg)_\varepsilon \in \mathcal{G}_\mathcal{S} (E),
\]
where $f_n\in E^{\widehat{\otimes}n}$ is associated with $(f_n^\varepsilon)_\varepsilon\in \mathcal{G}_\mathcal{S} (\mathbb{R}^3)^{\widehat{\otimes}n}$, and
 \[
 F(\zeta) \rightarrow  \mathbf{F}(\zeta) = \Bigg( \sum_{n=0}^\infty \langle :\, \zeta^{\otimes n} \, : , F_n^\varepsilon \rangle \Bigg)_\varepsilon \in\mathcal{G}_\tau (E)^* ,
 \]
where $F_n \in  (E^{\widehat{\otimes}  n})^\ast$  is associated with $(F^\varepsilon_n)_\varepsilon \in \mathcal{G}_\tau(\mathbb{R}^3)^{\widehat{\otimes}  n}$ and $ F_n^\varepsilon \in \mathcal{O}_M(\mathbb{R}^3)^{\widehat{\otimes}  n}$.
 %%%%%%%%%%%%%
\subsection{Compositions of generalised differential operators in Hida-Colombeau algebras}
We can now define the extended mappings
 \[
\mathbf{ D}_\eta \, : \, \mathcal{G}_\mathcal{S}(E)  \, \rightarrow \, \mathcal{G}_\mathcal{S} (E) \quad\mathrm{and}\quad \mathbf{ D}^*_\xi \, : \, \mathcal{G}_\tau (E)^* \, \rightarrow \, \mathcal{G}_\tau (E)^*\,,
 \] 
where
\[
\mathbf{ D}_\eta \,\mathbf{ f}(\zeta) = \sum_{n=0}^{\infty} n \langle :\zeta^{\otimes (n-1)}: \, , \eta^\varepsilon \, \hat{\otimes}_1 \,  f_n^\varepsilon \rangle\,, \qquad \zeta,\eta \in E^\ast ,
\]
and
 \[
 \mathbf{D}^*_\xi \, \mathbf{F}(\zeta) = \Bigg( \sum_{n=0}^\infty \langle :\, \zeta^{\otimes (n+1)} \, : , \xi^\varepsilon \hat{\otimes} F_n^\varepsilon \rangle \Bigg)_\varepsilon \, ,
 \]
 where $\eta, \xi \in S'(\mathbb{R}^3)$ are associated with $(\eta^\varepsilon)_\varepsilon,\, (\xi^\varepsilon)_\varepsilon \in \mathcal{G}_\tau(\mathbb{R}^3)$, respectively.

The composition
 \[
 \mathbf{D}_\eta \, [ \mathbf{D}^*_\xi  \mathbf{F}(\zeta)] = \Bigg( \sum_{n=0}^\infty  (n + 1)\langle :\, \zeta^{\otimes (n)} \, : ,  \eta^\varepsilon \hat{\otimes}_1(\xi^\varepsilon \hat{\otimes} F_n^\varepsilon) \rangle \Bigg)_\varepsilon \, ,
 \]
is well-defined generalised operator in $ \mathbf{L}(\mathcal{G}_\tau (E)^*, \mathcal{G}_\tau (E)^*)$, since for any $ \eta^\varepsilon, \xi^\varepsilon, F_{n_i}^\varepsilon\in \mathcal{O}_M(\mathbb{R}^3), n_i = 1 \dots n,$  expressions $\langle \eta^\varepsilon\,, \xi^\varepsilon\rangle$ and $\langle \eta^\varepsilon\,, F^\varepsilon_{n_i}\rangle$ are well-defined and $\eta^\varepsilon \hat{\otimes}_1(\xi^\varepsilon \hat{\otimes} F_n^\varepsilon)\in \mathcal{O}_M(\mathbb{R}^3)^{\widehat{\otimes}  n}$.

\subsection{Free real scalar fields as generalised operators in Hida-Colombeau algebras}
A Hida-Colombeau space representation of the free real scalar field can be written in the form
\[
\mathbf{\Phi}(t, \mathbf{x}) = \mathbf{\Phi}_-(t, \mathbf{x}) + \mathbf{\Phi}_+(t, \mathbf{x}) = \mathbf{D}_{z_1}(t, \mathbf{x}) + \mathbf{D}^*_{z_3}(t, \mathbf{x})\in\mathbf{L}(\mathcal{G}_\mathcal{S} (E), \mathcal{G}_\tau (E)^*) \, ,
\] 
and its canonical momentum field 
\[
\mathbf{\Pi}(t, \mathbf{x}) = \mathbf{\Pi}_-(t, \mathbf{x}) + \mathbf{\Pi}_+(t, \mathbf{x}) = i\,(-\mathbf{D}_{z_2}(t, \mathbf{x}) + \mathbf{D}^*_{z_4}(t, \mathbf{x}))\in\mathbf{L}(\mathcal{G}_\mathcal{S} (E), \mathcal{G}_\tau (E)^*) \,,
\]
where ordinary products $\mathbf{\Phi}^2, \mathbf{\Pi}^2, (\nabla \mathbf{\Phi})^2 \in \mathbf{L}(\mathcal{G}_\mathcal{S} (E), \mathcal{G}_\tau (E)^*)$ are well-defined generalised operators. Thus, the Hamiltonian of this field can be written as generalised operator in $ \mathbf{L}(\mathcal{G}_\mathcal{S} (E), \mathcal{G}_\tau (E)^*)$:
\begin{eqnarray*}
\lefteqn{
\mathbf{H}(\mathbf{\Phi})= \frac{1}{2}\int_{\mathbb{R}^3} [\mathbf{\Pi}^2 + (\nabla \mathbf{\Phi})^2 + m^2 \mathbf{\Phi}^2 ] d\mathbf{x}}\\ &&\quad
 = \frac{1}{2}\int_{\mathbb{R}^3} \omega(\mathbf{p})[ \partial_{\mathbf{p}}\partial_{\mathbf{p}}^* + \partial_{\mathbf{p}}^*\partial_{\mathbf{p}}] d\mathbf{p} = \frac{1}{2} \Big[\mathbf{D}_{\sqrt{\omega(\mathbf{p})}} \mathbf{D}^*_{\sqrt{\omega(\mathbf{p})}}\ + \mathbf{D}^*_{\sqrt{\omega(\mathbf{p})}} \mathbf{D}_{\sqrt{\omega(\mathbf{p})}}\Big]\,.
\end{eqnarray*}
\subsection{Self interacting scalar fields as generalised operators in Hida-Colombeau algebras}
While mathematically we could write the self interaction $V(\phi)$ of a scalar field as an polynomial of infinite order we note that renormalisable scalar fields can not have self interactions with terms that exceed $\phi^4$ \cite{Pesk}. Thus the Hida-Colombeau algebra we have developed so far is sufficient to provide a rigorous analysis of the these self interacting fields in terms of ordinary products. It is important to note that the physical Higgs field does have a $\phi^4$ self interaction term \cite{Pesk}.
\section{Discussion}
We have developed a rigorous axiomatic theory for self interacting scalar field that enables the fields to be multiplied as ordinary products rather than needing to use the (renormalised) Wick product.

Our axioms are those used by Sydney Coleman. We then use the Wiener-It$\hat{\mathrm{o}}$-Segal isomorphism to allow us to write the creation and annihilation operators for a many-body theory as generalised differential operators on the Hida distribution space, which simplifies the analysis.

However Hida distributions are subject to the Schwartz impossibility theorem and so in general do not commute. In 1985 Francois Colombeau showed that distributions could be embedded in factor-algebras of infinitely differentiable functions of moderate growth. We use generalised operators on simplified Hida-Colombeau algebras to represent scalar quantum fields and this allows us to write ordinary products of these fields. Hence dynamic operators such as the Hamiltonian can be written using these ordinary products.
 
\end{document}